\documentclass[a4paper,11pt]{article}
\pdfoutput=1
\usepackage{latexsym}
\usepackage{amsmath, amsthm, amssymb, bbm}
\usepackage[mathscr]{eucal}
\usepackage{fancyhdr}
\usepackage{tikz}
\usepackage{cite}
\usepackage{hyperref}
\usepackage{graphicx}

\newcommand{\wrt}{w.r.t.\ }
\newcommand{\cf}{cf.\ }

\newcommand{\ud}{\mathrm{d}}
\newcommand{\del}{\partial}

\newcommand{\R}{\mathbb{R}}

\newcommand{\order}{\mathcal{O}}

\newcommand{\sS}{\mathcal{S}}

\newcommand{\eff}{{\mathrm{eff}}}
\newcommand{\ph}{\varphi}

\newcommand{\bk}{{\mathrm{bulk}}}
\newcommand{\bd}{{\mathrm{boundary}}}

\begin{document}

\title{The excitation spectrum of rotating strings with masses at the ends}
\author{Jochen Zahn \\  \\ Fakult\"at f\"ur Physik, Universit\"at Wien, \\ Boltzmanngasse 5, 1090 Wien, Austria. \\ jochen.zahn@univie.ac.at}

\date{\today}

\maketitle

%\begin{flushright}
%UWThPh-2013-26
%\end{flushright}
%\hfill
%\vspace{0.7cm}
%\hfill
%\begin{center}
%{\LARGE The excitation spectrum of rotating strings with} \\
%\vspace{0.15cm}
%{\LARGE masses at the ends} \\
%\hfill
%\vspace{0.5cm}
%\hfill \\
%{\large Jochen Zahn \\  \vspace{0.3cm} Fakult\"at f\"ur Physik, Universit\"at Wien, \\ Boltzmanngasse 5, 1090 Wien, Austria \\ jochen.zahn@univie.ac.at \\
%\hfill \\
%\today \\}
%\hfill \\
%\end{center}

\begin{abstract}
We compute the spectrum of excitations of the rotating Nambu--Goto string with masses at the ends. We find interesting quasi-massless modes in the limit of slow rotation and comment on the nontrivial relation between world-sheet and target space energy.
\end{abstract}

\section{Introduction}

The Nambu--Goto string is a well-established phenomenological model for a vortex line connecting two quarks, \cf \cite{QWG04} and \cite{Teper09} for reviews on the situation in mesons and on the lattice.
In recent years, various aspects of this correspondence could be tested in lattice simulations, for example the corrections of order $\gamma^{-1} L^{-3}$ to the energy (where $L$ is the length of the string and $\gamma$ its tension), or the spectrum of the excitations of the string, at order $L^{-1}$.
For such studies, one usually considers the L\"uscher--Weisz string \cite{LuscherWeisz}, i.e., a string with fixed endpoints, corresponding to a string stretched between two stationary D0 branes.\footnote{The exact energy for (excitations of) this configuration was calculated by Arvis \cite{Arvis83}. We refer to \cite{AharonyKomargodski} for a recent critical reexamination of this result.} However, at least for light quarks, this does not seem to be a good approximation, as for reasonable values of the quark mass, the string tension, and the distance, one expects the quarks to rotate relativistically around each other. It is thus desirable to work with more realistic configurations.

Here, we study the Nambu--Goto string with masses at the ends, and compute the spectrum of excitations of the rigid solution, at first order in perturbation theory (corresponding to $\order(L^{-1})$). We find that, when moving away from the static limit, the degeneracy of the frequency of planar and scalar excitations (polarized in the plane of rotation or  orthogonal to it) is lifted. The degeneracy is restored in the relativistic limit, in which the ends of the string move at the speed of light. Furthermore, we identify interesting quasi-massless modes in the limit of slow rotation, which are not present in the L\"uscher--Weisz string.

The first treatment of the Nambu--Goto string with masses at the ends seems to have been the work of Chodos and Thorn \cite{ChodosThorn74}. They discussed the rigid rotating string solution, but did not study its excitations. Such a study was performed by Hadasz \cite{Hadasz99}, but for the case in which also a Gau{\ss}--Bonnet term is present. This leads to too many boundary conditions.\footnote{Concretely, a complex potential is used, and four boundary conditions are imposed at each end.}
In particular, the limit in which the coefficient of the Gau{\ss}--Bonnet term vanishes does not reduce to our case.

A non-relativistic approximation to the rotating string was studied by Nesterenko \cite{Nesterenko91}. We reproduce some of his results in the non-relativistic limit. However, our results differ in several aspects, in particular planar and scalar polarizations are not distinguished in \cite{Nesterenko91}.

The most recent study of the excitation spectrum of the rotating string with masses at the ends seems to have been the work of Baker and Steinke \cite{BakerSteinke02}. They use path integral techniques and obtain results for the spectrum of the excitations, which they study further in the limits of massless endpoints and one massless and one infinitely heavy endpoint. They use these to obtain Regge trajectories for the excited states. The main differences to our work are the following: We use perturbations that are normal to the background string, which simplifies the calculation considerably. We discuss the evolution of the spectrum over the whole range of parameters. In particular, we find interesting quasi-massless modes in the limit of a slowly rotating string. Finally, we have a different point of view regarding the significance of the spectrum, i.e., world-sheet energies, for the calculation of target space energies.

The article is structured as follows: In the next section, we introduce the setup and recall some basic results about the rotating rigid string. In Section~\ref{sec:Perturbations}, we consider perturbations of the rigid string and in particular compute their spectrum. For the limiting cases of a static and a massless string, this can be done analytically, while in the intermediate regime we use numerical methods.
In Section~\ref{sec:Energy} we argue that the relation of target space and world-sheet energies is more complicated than assumed, for example, in \cite{BakerSteinke02}. In particular, at the order usually considered, one already has to take the interaction into account.
We conclude with a summary and an outlook.

\section{Setup}

The Nambu--Goto action with boundary terms accounting for masses is given by\footnote{For simplicity, we here assume equal masses at the two ends.}
\begin{equation}
\label{eq:S}
 \sS = \sS_\bk + \sS_\bd = - \gamma \int_\Sigma \sqrt{- g} \ud^2 x - m \int_{\del \Sigma} \sqrt{- h} \ud x. 
\end{equation}
Here $\Sigma$ is the source space and $g$ the determinant of the induced metric, pulled back by the embedding $X: \Sigma \to \R^d$:
\[
 g_{\mu \nu} = \del_\mu X^a \eta_{ab} \del_\nu X^b.
\]
Here $\eta$ is the Minkowski metric on $\R^d$, with signature $(-, +, \dots, +)$. Likewise, $h$ is the induced metric pulled back to $\del \Sigma$.

On $\Sigma$, we choose coordinates $(\tau, \sigma) \in \R \times [-S, S]$. We determine the boundary conditions corresponding to the action \eqref{eq:S}, by the requirement that the boundary terms vanish. This leads to
\begin{equation}
\label{eq:BC}
 \left( \gamma g^{1 \nu} \del_\nu X^b \sqrt{- g} \pm m \del_0 \left( \frac{\del_0 X^b}{\sqrt{-  g_{00}}} \right) \right)|_{\sigma= \pm S} = 0.
\end{equation}

Let us now consider the rigid rotating string solution, which we parameterize as
\begin{equation}
\label{eq:Background}
 X = R ( \tau, \cos \tau \sin \sigma, \sin \tau \sin \sigma, 0 ),
\end{equation}
where $0$ stands for the $0$ of $\R^{d-3}$.
Here, we have to choose $S < \pi/2$. One easily computes
\begin{align*}
 g_{\mu \nu} & = R^2 \cos^2 \sigma \eta_{\mu \nu}, &
 \sqrt{-g} g^{\mu \nu} & = \eta^{\mu \nu},
\end{align*}
so that the boundary conditions are satisfied, provided that
\begin{equation}
\label{eq:BoundaryRelation}
 \frac{\gamma R}{m} = \frac{\tan S}{\cos S}.
\end{equation}
Two limits are of particular interest: The \emph{relativistic limit} $\gamma R/m \to \infty$, i.e., $S \to \pi/2$, in which the velocity of the ends of the strings approaches unity. And the \emph{static limit} $\gamma R/m \to 0$, while keeping $L = 2 \gamma R^2/m$ fixed, which corresponds to the string connecting two stationary D0 branes, separated by the distance $L$. 

The energy of a string configuration is given by \cite{Scherk75}
\begin{equation}
\label{eq:Energy}
 E = \int_{-S}^S \frac{\delta \sS_\bk}{\delta \del_0 X^0(0, \sigma)} \ud \sigma + \frac{\delta \sS_\bd}{\delta \del_0 X^0(0, -S)} + \frac{\delta \sS_\bd}{\delta \del_0 X^0(0, S)}.
\end{equation}
For the rotating string solution \eqref{eq:Background}, we compute, using \eqref{eq:BoundaryRelation}, the energy
\begin{equation*}
 E = 2 \gamma R S + 2 \gamma R \tfrac{1}{\tan S},
\end{equation*}
where the first term is due to the string and the second one due to the masses at the ends.
By comparison with the expression for the static string, we conclude that the \emph{effective length} of the string is given by
\[
 L_\eff = 2 R S.
\]
Note that this is greater than the length of the string measured in the laboratory frame, which is given by  $2 R \sin S$.

We note that the solution \eqref{eq:Background} breaks a couple of symmetries of the target Minkowski space. All spatial translations are broken, whereas time translation survives if accompanied by a rotation in the $1-2$ plane. Furthermore, all boosts are broken, as well as the rotations in the planes $1-2$, $1-i$, $2-i$, for $i \geq 3$. Below, we identify $d-2$ corresponding zero modes, consistently with the fact that locally we break $d-2$ translational symmetries \cite{LowManohar}. In the static limit, we will find quasi-massless modes for the remaining broken symmetries, with the exception of the boosts.\footnote{This is not surprising, as these can not be seen as infinitesimal for very late or early times.}

\section{Perturbations}
\label{sec:Perturbations}

Now we consider perturbations of the rotating string solution \eqref{eq:Background}, henceforth denoted by $\bar X$. That is, we write
\[
 X^a = \bar X^a + \ph^a,
\]
and expand the action in $\ph$. As we are perturbing around a classical solution, the component of first order in $\ph$ vanishes. The action at second order in $\ph$ defines the free part $\sS_0$ of the action. Due to the diffeomorphism invariance of the action, the equations of motion derived from $\sS_0$ are not hyperbolic. The physical degrees of freedom of the bulk part of the action are the normal perturbations \cite{BRZ}, i.e., those fulfilling
\[
 \del_\mu \bar X^a \eta_{ab} \ph^b = 0.
\]
The gauge condition we will adopt here is to set the longitudinal perturbations to $0$. This was apparently first proposed in \cite{Kleinert86}. Obviously, this gauge choice preserves the target space symmetries that are unbroken by the solution \eqref{eq:Background}.

In the following, we distinguish between the $d-3$ \emph{scalar} polarizations, $\ph^a = f^a_s e^a$, $a \geq 3$, which are orthogonal to the plane of rotation, and one \emph{planar} polarization, which may be written as
\begin{align}
\label{eq:planarMode}
 \ph^a & = f_p u^a, & u & = (\tan \sigma, - \sin \tau / \cos \sigma, \cos \tau / \cos \sigma ,0).
\end{align}
Due to the boundary condition, we have a supplementary physical degree of freedom at the boundary, namely radial perturbations. We may write them as
\begin{align}
\label{eq:radialMode}
 \ph^a & = f_r v^a, & v & = (0, \cos \tau, \sin \tau, 0),
\end{align}
where $f_r$ lives on the boundary, i.e., at $\sigma = \pm S$. This mode will enter the boundary conditions for the planar polarization. In terms of these modes, the free action $\sS_0$ can be written as
\begin{align}
\label{eq:S0}
 \sS_0 & = \frac{\gamma}{2} \int_\Sigma \left( {\dot f_p}^2 - {f'_p}^2 - \tfrac{2}{\cos^2 \sigma} f_p^2 + {\dot f_s}^2 - {f'_s}^2 \right) \ud^2 x  \\
 & + \frac{\gamma}{2} \frac{1}{\tan S} \int_{\del \Sigma} \left( {\dot f_p}^2 + {\dot f_r}^2 + {\dot f_s}^2 + \tfrac{1}{\cos^2 \sigma} f_p^2 + (1+2 \tan^2 \sigma) f_r^2 \right. \nonumber \\
 & \qquad \qquad \qquad \qquad \qquad \qquad \qquad \qquad \left. + \tfrac{2}{\cos \sigma} ( \dot f_p f_r - f_p \dot f_r )  \right) \ud x. \nonumber
\end{align}

\subsection{Equations of motion}

From the first term of \eqref{eq:S0}, we obtain the bulk equations of motion
\begin{align*}
 - \ddot f_s + f''_s & = 0, &
 - \ddot f_p + f_p'' - \tfrac{2}{\cos^2 \sigma} f_p & = 0.
\end{align*}
With the usual ansatz $f_{s/p}(\tau, \sigma) = e^{-ik \tau} f_{s/p,k}(\sigma)$, we have to solve the mode equations
\begin{align*}
 f''_{s,k} & = - k^2 f_{s,k}, &
 f_{p,k}'' - \tfrac{2}{\cos^2 \sigma} f_{p,k} & = - k^2 f_{p,k}.
\end{align*}

As we deal with a problem that is invariant under the reflection $\sigma \to - \sigma$, the solutions will be either even or odd. For the scalar polarization, the even/odd solutions are given by
\begin{align}
\label{eq:ModesScalar}
 f^+_{s, k} & = \cos k \sigma, &
 f^-_{s,k} & = \sin k \sigma.
\end{align}
The planar mode equation can be solved by choosing the coordinate $x = \sin \sigma$ and the ansatz $f(x) = (1-x^2)^{1/4} g(x)$, leading to
\[
 (1-x^2) g''(x) - 2 x g'(x) - \tfrac{9}{4} (1-x^2)^{-1} g(x) + (k^2 - \tfrac{1}{4}) g(x) = 0,
\]
which is the defining equation for Legendre functions \cite{AbramowitzStegun}. It follows that the general even/odd solutions of the planar mode equation are given by\footnote{Note that $Q^{3/2}_{1/2} = 0$ and $P^{3/2}_{1/2}$ is symmetric, so that $f_{p,k}^-$ as defined in \eqref{eq:f_p} degenerates for $k \to 1$. An antisymmetric solution for $k=1$ is given by
$f_{p, 1}^-(\sigma) = \frac{\sigma}{\cos \sigma} + \sin \sigma$.
Similarly, $f_{p,k}^+$ as defined in \eqref{eq:f_p} degenerates for $k=0$. The symmetric solution for $k=0$ may instead be written as
$f_{p, 0}^+(\sigma) = \sigma \tan \sigma + 1$.
}
\begin{align}
\label{eq:f_p}
 f^{\pm}_{p,k} & = \cos^{\frac{1}{2}} \sigma \left[ \frac{\sqrt{\pi}}{2} \sin ((k-\tfrac{1}{2}) \tfrac{\pi}{2}) \left( P_{k-\frac{1}{2}}^{\frac{3}{2}}(\sin \sigma) \pm P_{k-\frac{1}{2}}^{\frac{3}{2}}(- \sin \sigma) \right) \right. \\
 & \qquad \qquad \qquad \left. + \frac{1}{\sqrt{\pi}} \cos ((k-\tfrac{1}{2}) \tfrac{\pi}{2}) \left( Q_{k-\frac{1}{2}}^{\frac{3}{2}}(\sin \sigma) \pm Q_{k-\frac{1}{2}}^{\frac{3}{2}}(- \sin \sigma) \right) \right]. \nonumber
\end{align}

\subsection{Boundary conditions}

It remains to impose the correct boundary conditions. By restricting to even or odd solutions, it suffices to solve one of the boundary conditions. For the scalar polarization, we obtain, by expanding \eqref{eq:BC} to first order in $\ph$ and employing the relation \eqref{eq:BoundaryRelation},
\begin{equation}
\label{eq:BC_scalar}
 \tan S f_{s,k}'(S) - k^2 f_{s,k}(S) = 0.
\end{equation}
These are frequency dependent boundary conditions, similar to those derived for the nonrelativistic approximation in \cite{Nesterenko91}. For further occurrences of such boundary conditions, we refer to the references cited there.

For the even/odd modes in \eqref{eq:ModesScalar}, the boundary condition \eqref{eq:BC_scalar} reduces to the transcendental equations\footnote{\label{ft:Relation2BS}In \cite{BakerSteinke02}, these two equations correspond to the roots of the eigenvalues of the matrix ${\Gamma^{ij}_\theta}^{-1}$, \cf equation (10.6) there.}
\begin{align*}
 \tan S \tan k S & = - k, &
 \tan S & = k \tan k S.
\end{align*}
It follows straightforwardly that one always has the two solutions
\begin{align}
\label{eq:Scalar0Mode}
 f^+_{s,0} & = 1, &
 f^-_{s,1} & = \sin \sigma.
\end{align}
Furthermore, it is clear that no solutions with imaginary frequency exist, as $\cosh$ and $\sinh$ are both positive for positive argument. 
We note that for polarization in direction $e^i$, the $k_0$ mode (with real coefficient) corresponds to translations in direction $e^i$. Contrary to the $k_0$ mode, the $k_1$ mode is complex, due to the phase $e^{-i \tau}$. Hence, after multiplication with a complex coefficient $c^s_1$, one has to take the real part. Comparison with the background solution \eqref{eq:Background} shows that for polarization in the direction $e^i$, this corresponds to an infinitesimal rotation in the plane spanned by $e^1 \Re c^s_1 + e^2 \Im c^s_1$ and $e^i$. Below, we will see that this mode is quasi-massless in the static limit.

For the planar polarization, we have to take the radial mode \eqref{eq:radialMode} into account.
Upon contracting the first order expansion of \eqref{eq:BC} with $u$ and $v$, \cf \eqref{eq:planarMode} and \eqref{eq:radialMode}, we obtain
\begin{align*}
 \tan S f'_{p,k}(S) - \left( k^2 + \tfrac{1}{\cos^2 S} \right) f_{p,k}(S) - \tfrac{2 i k}{\cos S}  f_{r,k}(S) & = 0, \\
 \left( k^2 + 1 + 2 \tan^2 S \right) f_{r,k}(S) - \tfrac{2 i k}{\cos S} f_{p,k}(S) & = 0.
\end{align*}
Hence, the boundary condition for the planar polarization is
\begin{equation}
\label{eq:BC_planar}
 \tan S f'_{p,k}(S) - \left( k^2 + \tfrac{1}{\cos^2 S} - \tfrac{4 k^2}{(k^2 + 1 + 2 \tan^2 S) \cos^2 S} \right) f_{p,k}(S) = 0.
\end{equation}
In particular, for $k=0/1$, we obtain
\begin{align*}
 \tan S f'_{p,0}(S) - \tfrac{1}{\cos^2 S} f_{p,0}(S) & = 0, \\
 \tan S f'_{p,1}(S) - \left( \tfrac{1}{\cos^2 S} - 1 \right) f_{p,1}(S) & = 0.
\end{align*}
It is then straightforward to check that
\begin{align}
\label{eq:Planar0Mode}
f_{p,0}^- & = \tan \sigma, & f_{r,0}^- & = 0, \\
 f_{p,1}^+ & = - 1/\cos \sigma, & f_{r,1}^+ & = i \nonumber
\end{align}
always solve the boundary condition. Comparison of \eqref{eq:planarMode} and \eqref{eq:Background} shows that for small $S$, the $k_0$ mode corresponds to rotations in the $1-2$ plane, so we identify it with the Goldstone mode for this broken symmetry. As for the scalar $k_1$ mode, we multiply the planar $k_1$ mode with a complex coefficient $c^p_1$ and take the real part. For small $S$, this corresponds to a translation in the direction $e^2 \Re c^p_1 - e^1 \Im c^p_1$. Again, this mode is quasi-massless in the static limit. Hence, we have identified the (quasi-) Goldstone modes for spatial translations and rotations in the planes $1-2$, $1-i$, $2-i$ for $i \geq 3$.
%These correspond to rotations and translations in the plane of rotation.

To summarize, both for the scalar and the planar polarization, we have the two modes $k_0 = 0$, $k_1 = 1$. Note, however, that  the scalar $k_{0/1}$ mode is even/odd, while for the planar polarization this is converse.

\subsection{Solutions}

Let us analyze the solutions to the boundary conditions in the limiting cases $S \to 0$ and $S \to \pi/2$. In the first case, the scalar polarization boundary condition \eqref{eq:BC_scalar} reduces, for $k \gg 1$, to Dirichlet boundary conditions, so that we find $k = n \frac{\pi}{2 S}$, $n \in \{ 1, 2, \dots \}$. As we already have $k_0$ and $k_1$ modes, we define $k_n = (n-1) \frac{\pi}{2 S}$, $n \in \{ 2, 3, \dots \}$, where even/odd $n$ corresponds to an even/odd mode. Naively converting this to target space energies (but see Section~\ref{sec:Energy}), this corresponds to energies
\[
 \omega_n = \frac{k_n}{R} = (n-1) \frac{\pi}{L_\eff}, \qquad n \in \{ 2, 3, \dots \}.
\]
Note that the energy of the $k_1$ mode vanishes in the limit $S \to 0$, if expressed in natural units, as
\[
 \frac{L_\eff}{\pi} \omega_1 = \frac{2 S}{\pi}.
\]
%We note that for polarization in direction $e^i$, the $k_0$ mode corresponds to translations in direction $e^i$, whereas the $k_1$ mode corresponds, in the limit of a background string in direction $e^1$, to rotations in the $1-i$ plane.

For the planar polarization and $k \gg 1$, the boundary conditions \eqref{eq:BC_planar} also reduce to Dirichlet boundary conditions in the limit $S \to 0$. Using the expansions \cite[Eqns 8.10.7 \& 8.10.8]{AbramowitzStegun}, one sees that in the limit of large $k$, we have
\begin{align*}
 f^+_{p,k} & \sim \cos k \sigma, & f^-_{p,k} & \sim \sin k \sigma.
\end{align*}
Hence, we again find the solutions $k = n \frac{\pi}{2 S}$, $n \in \{ 1, 2, \dots \}$ in the limit $S \to 0$, where even/odd $n$ corresponds to an odd/even solution. We already identified the odd $k_0$ mode and the even $k_1$ mode. The first of the above modes is an even mode, so we are missing one odd mode. It turns out that it is given by $k_2 = \sqrt{3}$ in the limit $S \to 0$, for which the expression in brackets in \eqref{eq:BC_planar} vanishes. Hence, the remaining modes can be identified as $k_n = (n-2) \frac{\pi}{2 S}$, $n \in \{ 3, 4, \dots \}$. Note that, as for the $k_1$ mode in the scalar case, the energy of the planar modes $k_1$ and $k_2$ vanish in natural units in the limit $S \to 0$. Hence, apart from the supplementary quasi-massless modes, we find a degeneracy of the scalar and planar modes in the static limit. This is of course to be expected, as in this limit one can not distinguish scalar and planar polarization.

Let us now look at the limit $S \to \pi/2$. In the scalar case, the boundary condition \eqref{eq:BC_scalar} tends to Neumann boundary conditions, so we expect the solutions $k_n = n$, $n \in \{ 0, 1, \dots \}$, where even $n$ corresponds to an even mode and vice versa.

At first sight, the planar boundary condition \eqref{eq:BC_planar} seems to tend to Dirichlet boundary conditions in the limit $S \to \pi/2$, due to the $1/\cos^2 S$ dependence of the second term. However, the solutions $f_{p,k}^\pm$ generically behave as $(\sigma \mp \pi/2)^{-1}$ around $\pm \pi/2$, in which case both terms in \eqref{eq:BC_planar} behave as $(\sigma \mp \pi/2)^{-3}$. Hence, the situation is more complicated. By numerical analysis, one finds that in the limit $S \to \pi/2$, the $f_{p,k}^\pm$ solve the boundary conditions for odd/even $k$. This corresponds to $k_n = n, n \in \{ 0, 1, \dots \}$. Hence, in the relativistic limit, the scalar and planar modes also degenerate, but now even scalar modes degenerate with odd planar modes and vice versa. Hence, we expect that for $0 < S < \pi/2$, the degeneracy is lifted. Table~\ref{tab:Modes} summarizes the properties of the modes in the limits $S \to 0$ and $S \to \pi/2$.

\begin{table}
\begin{tabular}{|c|c|c|c|c|c|c|}
\hline
 & \multicolumn{3}{|c|}{scalar} & \multicolumn{3}{|c|}{planar} \\
\hline
$n$ & $k_n$ static & $k_n$ rel.\ & parity & $k_n$ static & $k_n$ rel.\ & parity \\
\hline
0 & 0 & 0 & $+$ & 0 & 0 & $-$ \\
1 & 1 & 1 & $-$ & 1 & 1 & $+$ \\
2 & $\frac{\pi}{2 S}$ & 2 & $+$ & $\sqrt{3}$ & 2 & $-$ \\
$n > 2$ & $(n-1) \frac{\pi}{2 S}$ & $n$ & $(-1)^n$ & $(n-2) \frac{\pi}{2 S}$ & $n$ & $(-1)^{n-1}$ \\
\hline
\end{tabular}
\caption{The parity and the frequency of the scalar and planar modes in the static ($S \to 0$) and relativistic ($S \to \pi/2$) limit.}
\label{tab:Modes}
\end{table}

\begin{figure}
\centering
\includegraphics[width=\textwidth]{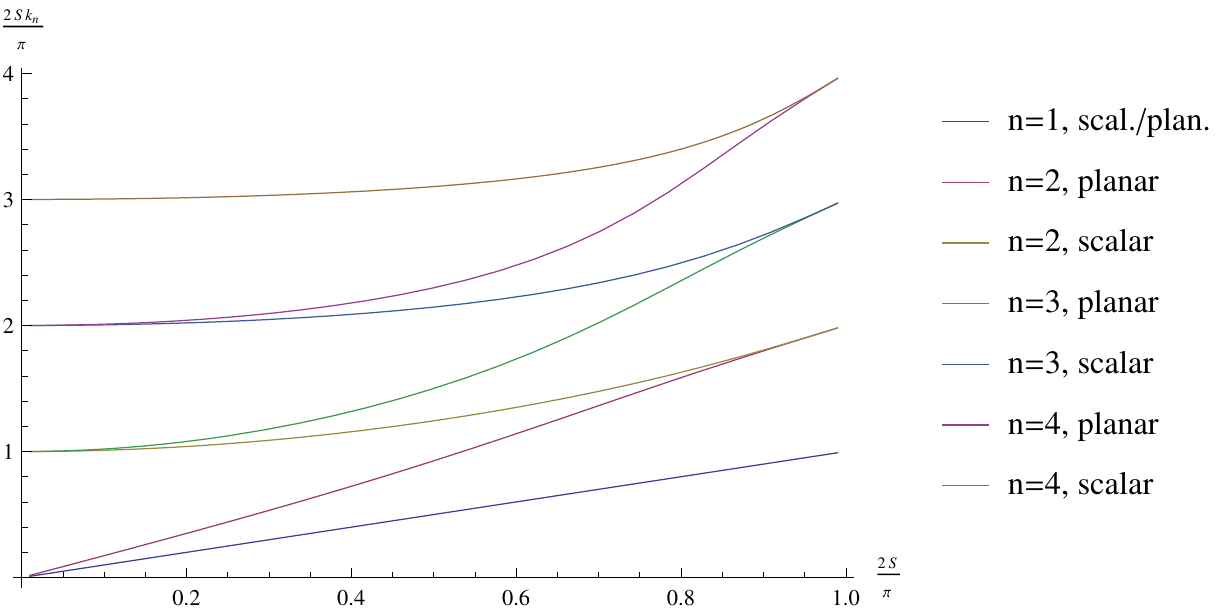}
\caption{The frequency of the first few modes as a function of $S$. Note that both the scalar and the planar $n=0$ mode have vanishing frequency.}
\label{fig:Spectrum}
\end{figure}

For $0 < S < \pi/2$, the frequencies of the modes can be found numerically. Figure~\ref{fig:Spectrum} shows these as a function of $S$. As expected, the degeneracy of the scalar and planar modes is lifted. In the limit of a slowly rotating string, there are three quasi massless modes, the two $n=1$ modes and the planar $n=2$ mode. To the best of our knowledge, these have not been discussed before. The first two have a geometric interpretation, namely rotations in the scalar and translations in the planar case. The significance of the planar $n=2$ mode is presently unclear, but it could be relevant for stability, \cf the discussion of the ``roughening transition'' in \cite{Luscher81}, for example. 

\section{World-sheet and target space energy}
\label{sec:Energy}

In order to interpret our results, let us first note that the free Hamiltonian corresponding to $\sS_0$ is given by
\begin{align}
 H(\tau) & = \frac{\gamma}{2} \int \left(  {\dot f_p}^2 + {f'_p}^2 + \tfrac{2}{\cos^2 \sigma} f_p^2 + {\dot f_s}^2 + {f'_s}^2 \right) \ud \sigma \nonumber \\
\label{eq:Hamiltonian}
 & + \frac{\gamma}{2} \frac{1}{\tan S} \left( \dot f_p^2| + \dot f_r^2| - \tfrac{1}{\cos^2 S} f_p^2| - (1 + 2 \tan^2 S) f_r^2| + \dot f_s^2| \right),
\end{align}
where in the first line the $f$'s are evaluated at $(\tau, \sigma)$, and in the second line $f^2| = f(\tau, S)^2 + f(\tau, -S)^2$. Denoting by $k_n^{s/p}$ the $n$th scalar/planar mode frequency, we may normalize the modes,
\begin{align*}
 \phi^{s}_n(\sigma) & = c^s_n f_{s,k^s_n}^{(-1)^n}(\sigma), &
 \phi^{p/r}_n(\sigma) & = c^p_n f_{p/r,k^p_n}^{(-1)^{n-1}}(\sigma),
\end{align*}
such that, for $m, n \neq 0$,
\begin{align}
\label{eq:Orthogonal_Scalar}
  \int \phi^s_m \phi^s_n \ud \sigma + \tfrac{1}{\tan S} \phi^s_m \phi^s_n| & = \delta_{m n}, \\
  \int \phi^p_m \phi^p_n \ud \sigma + \tfrac{1}{\tan S} \phi^p_m \phi^p_n| + \tfrac{1+ 2 \tan^2 S}{\tan S} \tfrac{1}{k^p_m k^p_n} \phi^r_m \phi^r_n| & = \delta_{mn}, \nonumber
\end{align} 
where $\phi \psi|$ stands for $\phi (S) \psi(S) + \phi(-S) \psi(-S)$. Furthermore, one sets, for positive $n$,
\begin{align*}
 k^{s/p}_{-n} & = - k^{s/p}_n, &
 \phi^{s/p}_{-n}(\sigma) & = \phi^{s/p}_{n}(\sigma), &
 \phi^{r}_{-n}(\sigma) & = - \phi^{r}_{n}(\sigma).
\end{align*}
For $n=0$, we choose the normalization constant $c^{s/p}_0 = 1$, and note that, for $m \neq 0$,
\begin{align}
 \int \phi^s_m \phi^s_0 \ud \sigma + \tfrac{1}{\tan S} \phi^s_m \phi^s_0| & = 0, \nonumber \\
\label{eq:Orth0_Planar}
 k_m \int \phi^p_m \phi^p_0 \ud \sigma + \tfrac{k_m}{\tan S} \phi^p_m \phi^p_0| + \tfrac{2 i}{\sin S} \phi^r_m \phi_0^p| & = 0.
\end{align}
We pursue a canonical quantization approach based on the modes, and write, recalling the zero modes in \eqref{eq:Scalar0Mode} and \eqref{eq:Planar0Mode},
\begin{align}
\label{eq:f_s_modes}
 f_s^i & = c_s x^i + d_s \tau p^i + i \frac{1}{\sqrt{2 \gamma}} \sum_{n \neq 0} e^{- i k^s_n \tau} \frac{\alpha^i_n}{k^s_n} \phi^s_n(\sigma), \\
\begin{pmatrix} f_p \\ f_r \end{pmatrix} & = ( c_p \theta + d_p \tau q) \begin{pmatrix} \tan \sigma \\ 0 \end{pmatrix} + i \frac{1}{\sqrt{2 \gamma}} \sum_{n \neq 0} e^{- i k^p_n \tau} \frac{\alpha^p_n}{k^p_n} \begin{pmatrix} \phi^p_n(\sigma) \\ \phi^r_n(\sigma) \end{pmatrix}, \nonumber
\end{align}
where $i \in \{ 3, \dots, d-1 \}$, $\alpha_n^* = \alpha_{-n}$, and
\begin{align*}
 c_s & = \sqrt{\tfrac{1}{2 \gamma (S + 1/\tan S)}}, &
 d_s & = \sqrt{\tfrac{1}{2 \gamma (S + 1/\tan S)}}, \\
 c_p & = - \sqrt{\tfrac{2 \tan S - S}{2 \gamma}} \tfrac{1}{2 \cos S \sin S/(1+\sin^2 S) + S}, &
 d_p & = \sqrt{\tfrac{1}{2 \gamma (2 \tan S - S)}}.
\end{align*}
The normalization is chosen such that, upon quantization, the canonical commutation relations between $f_{s/p/r}$ and their associated momenta $\pi_{s/p/r}$ imply
\begin{align}
 [x^i, p^j] & = i \hbar \delta^{ij}, &
 [\theta, q] & = i \hbar, \nonumber \\
\label{eq:CommRel}
 [ \alpha^i_n, \alpha^j_m ] & = \hbar \delta^{ij} k_n^s \delta_{n, -m}, &
 [ \alpha^p_n, \alpha^p_m ] & = \hbar k_n^p \delta_{n, -m}.
\end{align}
The free Hamiltonian can then be written as\footnote{As we argue below that this observable is not related to the target space energy, we refrain from calculating the vacuum energy here.}
\begin{equation}
\label{eq:HamiltonianLadder}
 H(0) = \tfrac{1}{2} \left( p^2 + q^2 \right) + \sum_{n=1}^\infty \left( {\alpha^i_n}^* \alpha^i_n + {\alpha^p_n}^* \alpha^p_n \right).
\end{equation}
Hence, due to the commutation relations \eqref{eq:CommRel}, the world-sheet energy, i.e., the energy \wrt to the time $\tau$ on the world-sheet, of the first excited state of the scalar/planar $n$th mode is $E_n^{s/p} = \hbar k^{s/p}_n$.

There now seem to be two possibilities to define the target space energy. The first one is to take \eqref{eq:Energy} and expand it in $\ph$. The second one\footnote{This possibility is chosen in \cite{Nesterenko91} and \cite{BakerSteinke02}. However, the world-sheet and the target space time evolution seem to be simply equated there, without taking the rotation into account.} is to relate the target space energy to the world-sheet energy. A translation in $\tau$ corresponds in target space to a translation in time and a rotation in the 1-2 plane. By the choice of the classical solution \eqref{eq:Background}, the target space time translation invariance is broken to discrete time translations $x^0 \to x^0 + 2 \pi R$, corresponding to translations $\tau \to \tau + 2 \pi$ on the world-sheet. The broken time translation invariance corresponds to considering energies modulo $1/R$. This reasoning suggests that, up to this ambiguity, the target space energy is given by the world-sheet energy divided by $R$. This would mean that, in the massless limit $S \to \pi/2$, all target space energies are equivalent to $0$ modulo $1/R$.

However, there already is a definition of the target space energy which we used to compute the energy of the classical solution, namely \eqref{eq:Energy}. Thus, it seems more natural to consider the expansion of \eqref{eq:Energy} in $\ph$. The first non-trivial term in this expansion is at first order, concretely,
\begin{equation}
\label{eq:TargetSpaceEnergy}
 E^{(1)} = \gamma \left[ \int \dot f_p \tan \sigma \ud \sigma + \dot f_p| + \tfrac{2}{\cos S} f_r| \right] = d_p^{-1} q,
\end{equation}
where we used \eqref{eq:Orth0_Planar} and the same notation as in \eqref{eq:Hamiltonian}. Two facts are noteworthy: First, the world-sheet Hamiltonian \eqref{eq:Hamiltonian} does not have a linear term, so there does not seem to be any relation to the world-sheet energy. Second, as the target space energy does have a contribution at first order in $\ph$, the computation of higher order contributions does have to take the interaction into account. To see what happens if the interaction is neglected, consider the contribution to the target space energy from the scalar polarizations. One obtains, at second order in $\ph$,
\begin{equation}
\label{eq:TargetSpaceEnergyScalar}
 E^{(2)}_s = \frac{\gamma}{2R} \left[ \int_{-S}^S \tfrac{1}{\cos^2 \sigma} \left( {\dot f_s}^2 + {f'_s}^2 \right) \ud \sigma + \tfrac{1}{\tan S \cos^2 S} {\dot f_s}^2| \right].
\end{equation}
This corresponds to the scalar contribution to the Hamiltonian \eqref{eq:Hamiltonian}, except for the factor $1/R$ and the weight $1/\cos^2 \sigma$. Due to this weight, upon replacing $f_s$ by the expansion \eqref{eq:f_s_modes}, we can no longer use the orthogonality relations \eqref{eq:Orthogonal_Scalar} to conclude that only the combinations ${\alpha^i_n}^* \alpha^i_n$ and $\alpha^i_n {\alpha^i_n}^*$ occur. Put differently, this quantity is not conserved. Of course the full target space energy \eqref{eq:Energy} is conserved, so in a perturbative expansion, one has conservation at each order. The correct scalar contribution to $E$ at second order has to take into account the contribution of the interaction terms $\ph_{p/r} \ph_s^2$ to $E^{(1)}$. Note that $E^{(1)}$ itself is conserved, but only under the free time-evolution.

To conclude, let us discuss the static limit. Then the rotation is negligible and one would expect that one can indeed identify world-sheet and target space energies by scaling with $R$. Let us see how this arises in our expression for the target space energy. In the limit $S \to 0$, $d_p$ diverges, so the first order contribution \eqref{eq:TargetSpaceEnergy} tends to $0$. Furthermore, the weight $1/\cos^2 \sigma$ in \eqref{eq:TargetSpaceEnergyScalar} tends to $1$, so one obtains the scalar contribution to the world-sheet Hamiltonian \eqref{eq:Hamiltonian}, up to the factor $1/R$.

\section{Conclusion}

We computed the spectrum of excitations of the rotating Nambu--Goto string, with masses at the ends. We distinguished between scalar and planar polarizations and showed that the degeneracy of these in the static limit (corresponding to the L\"uscher--Weisz string) is lifted for finite rotation speed. In the massless limit, the degeneracy is again restored. We identified interesting quasi-massless modes in the limit of the slowly rotating string.
We discussed two possible definitions of the target space energy, a naive one which relates world-sheet to target space energies, and one that is based on the perturbative expansion of the standard expression for the target space energy. We also showed that the two notions coincide in the static limit.

An interesting further goal would be to obtain quantum corrections to the classical Regge trajectories. These were computed in \cite{BakerSteinke02}, but using the naive identification of world-sheet and target space energies. The next logical step would thus be to compute the target space energy at second order in $\ph$. As discussed in Section~\ref{sec:Energy}, this requires to take the interaction into account.

\subsection*{Acknowledgments}
I would like to thank Christian K\"ohler, Jan Schlemmer, and Harold Steinacker for helpful discussions and remarks.
This work was supported by the Austrian Science Fund (FWF) under the contract P24713.

%\bibliography{../../mybib}{}
%%%\bibliographystyle{../halpha}
%%\bibliographystyle{../h-elsevier}
%\bibliographystyle{../../h-elsevier_new}

\end{document}